\documentclass{emulateapj}
\usepackage{apjfonts}

\shortauthors{Winn et al.~2009}
\shorttitle{Spin-Orbit Misalignment in HD~80606}

\begin{document}

% ------------------------------------------------------------------------
% New commands
%
\def\ltsima{$\; \buildrel < \over \sim \;$}
\def\lsim{\lower.5ex\hbox{\ltsima}}
\def\gtsima{$\; \buildrel > \over \sim \;$}
\def\gsim{\lower.5ex\hbox{\gtsima}}
                                                                                          
% -------------------------------------------------------------------------
%

\bibliographystyle{apj}

\title{
The Transit Ingress and the Tilted Orbit of the \\
Extraordinarily Eccentric Exoplanet HD~80606\lowercase{b}
}

\author{
Joshua N.\ Winn\altaffilmark{1},
Andrew W.\ Howard\altaffilmark{2,3},
John Asher Johnson\altaffilmark{4,5},
Geoffrey W.\ Marcy\altaffilmark{2},
J.~Zachary Gazak\altaffilmark{4},\\
Donn Starkey\altaffilmark{6},
Eric B.\ Ford\altaffilmark{7},
Knicole D.\ Col\'{o}n\altaffilmark{7},
Francisco Reyes\altaffilmark{7},
Lisa Nortmann\altaffilmark{8},
Stefan Dreizler\altaffilmark{8},\\
Stephen Odewahn\altaffilmark{9},
William F.\ Welsh\altaffilmark{10},
Shimonee Kadakia\altaffilmark{10},
Robert J.\ Vanderbei\altaffilmark{11},
Elisabeth R.\ Adams\altaffilmark{12},\\
Matthew Lockhart\altaffilmark{12},
Ian J.\ Crossfield\altaffilmark{13},
Jeff A.\ Valenti\altaffilmark{14},
Ronald Dantowitz\altaffilmark{15},
Joshua A.\ Carter\altaffilmark{1}
}

\submitted{Accepted for publication in the Astrophysical Journal (2009~August~18)}

\altaffiltext{1}{Department of Physics, and Kavli Institute for
  Astrophysics and Space Research, Massachusetts Institute of
  Technology, Cambridge, MA 02139}

\altaffiltext{2}{Department of Astronomy, University of California,
  Mail Code 3411, Berkeley, CA 94720}

\altaffiltext{3}{Townes Postdoctoral Fellow, Space Sciences
  Laboratory, University of California, Berkeley, CA 94720}

\altaffiltext{4}{Institute for Astronomy, University of Hawaii,
  Honolulu, HI 96822}

\altaffiltext{5}{NSF Astronomy and Astrophysics Postdoctoral Fellow}

\altaffiltext{6}{DeKalb Observatory H63, Auburn, IN 46706}

\altaffiltext{7}{Department of Astronomy, University of Florida, 211
  Bryant Space Center, P.O.\ Box 112055, Gainesville, FL 32611}

\altaffiltext{8}{Institute for Astrophysics, Friedrich-Hund-Platz 1,
  D-37077 G\"ottingen, Germany}

\altaffiltext{9}{McDonald Observatory, University of Texas, Austin, TX
  78712}

\altaffiltext{10}{Department of Astronomy, San Diego State University,
  5500 Campanile Drive, San Diego, CA 92182}

\altaffiltext{11}{Department of Operations Research and Financial
  Engineering, Princeton University, Princeton, NJ 08544}

\altaffiltext{12}{Department of Earth, Atmospheric, and Planetary
  Science, Massachusetts Institute of Technology, Cambridge, MA 02139}

\altaffiltext{13}{Department of Astronomy, University of California,
  430 Portola Plaza, Box 951547, Los Angeles, CA 90095}

\altaffiltext{14}{Space Telescope Science Institute, 3700 San Martin
  Dr., Baltimore, MD 21218}

\altaffiltext{15}{Clay Center Observatory, Dexter and Southfield
  Schools, 20 Newton St., Brookline, MA 02445}

\begin{abstract}

  We present the results of a transcontinental campaign to observe the
  2009~June~5 transit of the exoplanet HD~80606b. We report the first
  detection of the transit ingress, revealing the transit duration to
  be $11.64\pm 0.25$~hr and allowing more robust determinations of the
  system parameters. Keck spectra obtained at midtransit exhibit an
  anomalous blueshift, giving definitive evidence that the stellar
  spin axis and planetary orbital axis are misaligned. The Keck data
  show that the projected spin-orbit angle $\lambda$ is between
  32--87~deg with 68.3\% confidence and between 14--142~deg with
  99.73\% confidence. Thus the orbit of this planet is not only highly
  eccentric ($e=0.93$) but is also tilted away from the equatorial
  plane of its parent star. A large tilt had been predicted, based on
  the idea that the planet's eccentric orbit was caused by the Kozai
  mechanism. Independently of the theory, it is noteworthy that all 3
  exoplanetary systems with known spin-orbit misalignments have
  massive planets on eccentric orbits, suggesting that those systems
  migrate through a different channel than lower-mass planets on
  circular orbits.

\end{abstract}


\section{Introduction}

Discovered by Naef et al.~(2001), HD~80606b is a giant planet of
approximately 4 Jupiter masses whose orbit carries it within 7 stellar
radii of its parent star. Yet it is no ordinary ``hot Jupiter'': the
other end of the planet's 111-day orbit is about 30 times further away
from the star. With an orbital eccentricity of 0.93, HD~80606b
presents an extreme example of the ``eccentric exoplanet'' problem:
the observation that exoplanets often have eccentric orbits, despite
the 20th-century expectation that more circular orbits would be common
(Lissauer 1995).

Wu and Murray (2003) proposed that HD~80606b formed on a wide circular
orbit that was subsequently shrunk and elongated by a combination of
the Kozai (1962) effect and tidal friction. In this scenario, the
gravitational perturbation from the companion star HD~80607 excites
large-amplitude oscillations of the planet's orbital eccentricity and
inclination. During high-eccentricity phases, tidal friction drains
the orbital energy and shrinks the orbit until the oscillations cease
due to competing perturbations arising from stellar asphericity or
general relativity. Fabrycky \& Tremaine~(2007) noted that a probable
consequence of this scenario is that the star-planet orbit was left
tilted with respect to its original orbital plane, which was
presumably aligned with the stellar equator. Hence, a demonstration
that the planetary orbital axis and stellar spin axis are misaligned
would be supporting evidence for the Kozai scenario.

For a transiting planet, it is possible to measure the angle between
the sky projection of those two axes through observations of the
Rossiter-McLaughlin (RM) effect, a distortion of spectral lines
resulting from the partial eclipse of the rotating stellar surface
(Rossiter~1924, McLaughlin 1924, Queloz et al.~2000; see Fabrycky \&
Winn 2009 for a recent summary of results). In a series of fortunate
events, it recently became known that the orbit of HD~80606b is viewed
close enough to edge-on to exhibit transits and thereby permit RM
observations. First, Laughlin et al.~(2009) detected an occultation of
the planet by the star, an event that is visible from only 15\% of the
sight-lines to HD~80606. Then, three groups detected a transit (Moutou
et al.~2009, Fossey et al.~2009, Garcia-Melendo \& McCullough 2009),
which was predicted to occur with only 15\% probability even after
taking into account the occurrence of occultations.

All three groups detected the transit egress, but not the ingress. The
lack of information about the ingress, and hence the transit duration,
hampered previous determinations of this system's parameters. In
particular, Moutou et al.~(2009) gathered radial-velocity data
bracketing the transit egress that displays the RM effect, but due to
the unknown transit duration it was not immediately clear whether
meaningful constraints could be placed on the angle $\lambda$ between
the sky projections of the stellar spin axis and the orbital
axis. Pont et al.~(2009) concluded that $\lambda$ is nonzero based on
a Bayesian analysis of the available data, but their results were
sensitive to prior assumptions regarding the stellar mean density, the
stellar rotation rate, and the treatment of correlated noise, and were
therefore not as robust as desired.\footnote{Gillon (2009) has
  submitted for publication a similar analysis of the same data, with
  similar results.}

We report here on a campaign to observe the photometric transit
ingress of UT~2009~June~5, and to measure more precise radial
velocities during the transit. We also present and analyze data that
have been accumulated by the California Planet Search over the 8~years
since the planet's discovery. Our observations and data reduction are
presented in \S~2, our analysis is described in \S~3, and the results
are summarized and discussed in \S~4.

\section{Observations}

\subsection{Photometry}
\label{subsec:photometry}

The ingress was expected to begin between UT~23:00 June 4 and
06:00~June~5, and to last 4--5~hr. However, in June, HD~80606 is only
observable from a given site for a few hours (at most) following
evening twilight. To overcome this obstacle we organized a
transcontinental\footnote{We use this term with apologies to Hawaii,
  which is not even on the same tectonic plate as the other sites.}
campaign, with observers in Massachusetts, New Jersey, Florida,
Indiana, Texas, Arizona, California, and Hawaii.

On the transit night, each observer obtained a series of images of
HD~80606 and its neighbor HD~80607. In most cases, we used only a
small subraster of the CCD encompassing both stars, and defocused the
telescope, both of which allow an increase in the fraction of time
spent collecting photons as opposed to reading out the CCD. Defocusing
also has the salutary effects of averaging over pixel-to-pixel
sensitivity variations, and reducing the impact of natural seeing
variations on the shape of the stellar images.

Each observer also gathered images on at least one other night when
the transit was not occurring, to establish the baseline flux ratio
between HD~80606 and HD~80607 with the same equipment, bandpass, and
range of airmass as on the transit night. Details about each site are
given below. (Observations were also attempted from Brookline, MA;
Princeton, NJ; Lick Observatory, CA; and Winer Observatory, AZ; but no
useful data were obtained at those sites due to poor weather.) In what
follows, the dates are UT dates, i.e., ``June 5'' refers to the
transit night of June 4-5 in U.S.\ time zones.

{\it George R.\ Wallace Jr.\ Astrophysical Observatory, Westford, MA.}
Thick clouds on the transit night prevented any useful data from being
obtained. However, out-of-transit data in the Cousins $R$ band were
obtained on June~3 using a 0.41~m telescope equipped with a POETS
camera (Souza et al.~2006), and a 0.36~m telescope equipped with an
SBIG STL-1001E CCD camera.

{\it Rosemary Hill Observatory, Bronson, FL.} We observed in the Sloan
$i$ band using the 0.76~m Tinsley telescope and SBIG ST-402ME CCD
camera. Conditions were partly cloudy on the transit night, leading to
several interruptions in the time series. Control data were also
obtained on June~11.

{\it De Kalb Observatory, Auburn, IN.} We used a 0.41~m $f$/8.5
Ritchey-Chretien telescope with an SBIG ST10-XME CCD camera. Data were
obtained in the Cousins $R$ band on May~30 and on the transit
night. Conditions were clear on both nights.

{\it McDonald Observatory, Fort Davis, TX.} Two telescopes were used:
the McDonald 0.8m telescope and its Loral 2048$^2$ prime focus CCD
camera with an $R_C$ filter; and the MONET-North\footnote{MONET stands
  for MOnitoring NEtwork of Telescopes; see Hessman~(2001).} 1.2m
telescope with an Alta Apogee E47 CCD camera and SDSS $r$ filter,
controlled remotely from G\"ottingen, Germany. Conditions were partly
cloudy on the transit night, shortening the interval of observations
and causing several interruptions. Control data were obtained with the
McDonald 0.8m telescope on June~11, and with the MONET-North 1.2m
telescope on May~31 and June~4.

{\it Fred L.\ Whipple Observatory, Mt.\ Hopkins, AZ.} We used the
48~in (1.2m) telescope and Keplercam, a 4096$^2$ Fairchild CCD camera.
Cloud cover prevented observations on the transit night, but
out-of-transit data were obtained on June~6 in the Sloan $riz$ bands.
Out-of-transit data in the $i$ band were also obtained on Feb.~13 and
14, 2009.

{\it Mount Laguna Observatory, San Diego, CA.} We observed in the
Sloan $r$ band with the 1.0m telescope and 2048$^2$ CCD camera.
Conditions were humid and cloudy. The target star was observable
through a thin strip of clear sky for several hours after evening
twilight. Out-of-transit data were obtained on June~8.

{\it Mauna Kea Observatory, HI.} We used the University of Hawaii 2.2m
telescope and the Orthogonal Parallel Transfer Imaging Camera (OPTIC;
Tonry et al.~1997). Instead of defocusing, we used the charge-shifting
capability of OPTIC to spread the starlight into squares 40 pixels
($5\farcs 4$) on a side (Howell et al.~2003). We observed with a
custom ``narrow $z$'' filter defining a bandpass centered at 850~nm
with a full width at half-maximum of 40~nm. Out-of-transit data were
obtained on June~4.

Reduction of the CCD images from each observatory involved standard
procedures for bias subtraction, flat-field division, and aperture
photometry. The flux of HD~80606 was divided by that of HD~80607, and
the results were averaged into 10~min bins. This degree of binning was
acceptable because it sampled the ingress duration with $\approx$15
points. We estimated the uncertainty in each binned point as the
standard deviation of the mean of all the individual data points
contributing to the bin (ranging in number from 8 to 63 depending on
the telescope). We further imposed a minimum uncertainty of 0.001 per
10~min binned point, to avoid overweighting any particular point and
out of general caution about time-correlated noise that often afflicts
photometric data (Pont et al.~2006). Fig.~1 shows the time series of
the flux ratio based on the data from the transit night of June~5, as
well as the out-of-transit flux ratio derived with the same telescope.

%%%%%%%%%%%%%%%%%%%%%%%%%%%%%%%%%%%%%%%%%%%%%%%%%%%%%%%%%%%%%%%%%

\tabletypesize{\scriptsize}

\begin{deluxetable}{clccc}

\tabletypesize{\scriptsize}
\tablecaption{Out-of-transit flux ratio between HD~80606 and HD~80607\label{tbl:ratio}}
\tablewidth{0pt}

\tablehead{
\colhead{No.} &
\colhead{Observatory/Telescope} &
\colhead{Date} &
\colhead{Band} &
\colhead{Flux Ratio}
}

\startdata
$1$ & University of London 0.35m & 2009~Feb~14 & $R_C$ & $1.12796\pm 0.00023$ \\
$2$ & De Kalb 0.41m              & 2009~May~30 & $R_C$ & $1.12305\pm 0.00110$ \\
$3$ & Wallace 0.41m              & 2009~Jun~03 & $R_C$ & $1.12859\pm 0.00046$ \\
$4$ & Wallace 0.36m              & 2009~Jun~03 & $R_C$ & $1.12582\pm 0.00057$ \\
$5$ & McDonald 0.8m              & 2009~Jun~11 & $R_C$ & $1.12230\pm 0.00130$ \\
$6$ & MONET-North 1.2m           & 2009~May~31 & $r$   & $1.12281\pm 0.00062$ \\
$7$ & MONET-North 1.2m           & 2009~Jun~04 & $r$   & $1.12240\pm 0.00060$ \\
$8$ & Mt.\ Laguna 1m             & 2009~Jun~08 & $r$   & $1.12592\pm 0.00290$ \\
$9$ & Whipple 1.2m               & 2009~Jun~06 & $r$   & $1.12565\pm 0.00320$ \\
$10$ & Rosemary Hill 0.76m        & 2009~Jun~11 & $i$   & $1.12072\pm 0.00041$ \\
$11$ & Whipple 1.2m               & 2009~Jun~06 & $i$   & $1.11927\pm 0.00510$ \\
$12$ & Whipple 1.2m               & 2009~Feb~13 & $i$   & $1.11758\pm 0.00085$ \\
$13$ & Whipple 1.2m               & 2009~Feb~14 & $i$   & $1.11814\pm 0.00066$ \\
$14$ & Mauna Kea UH~2.2m          & 2009~Jun~04 & $z$\tablenotemark{a} & $1.11635\pm 0.00021$ \\
$15$ & Whipple 1.2m               & 2009~Jun~06 & $z$   & $1.11584\pm 0.00047$
\enddata

\tablenotetext{a}{A custom ``narrow $z$'' bandpass, centered at 850~nm
  with a full width at half-maximum of 40~nm.}

\tablecomments{Based on data from our campaign, except for the data
  from the University of London Observatory which was kindly provided
  by Fossey et al.~(2009). The quoted uncertainties represent only the
  ``statistical error,'' defined as the standard error of the mean of
  the flux ratios derived from all the images.}

\end{deluxetable}

%%%%%%%%%%%%%%%%%%%%%%%%%%%%%%%%%%%%%%%%%%%%%%%%%%%%%%%%%%%%%%%%%

\begin{figure*}[t]
\begin{center}
  \leavevmode
\hbox{%
  \epsfxsize=6in
  \epsffile{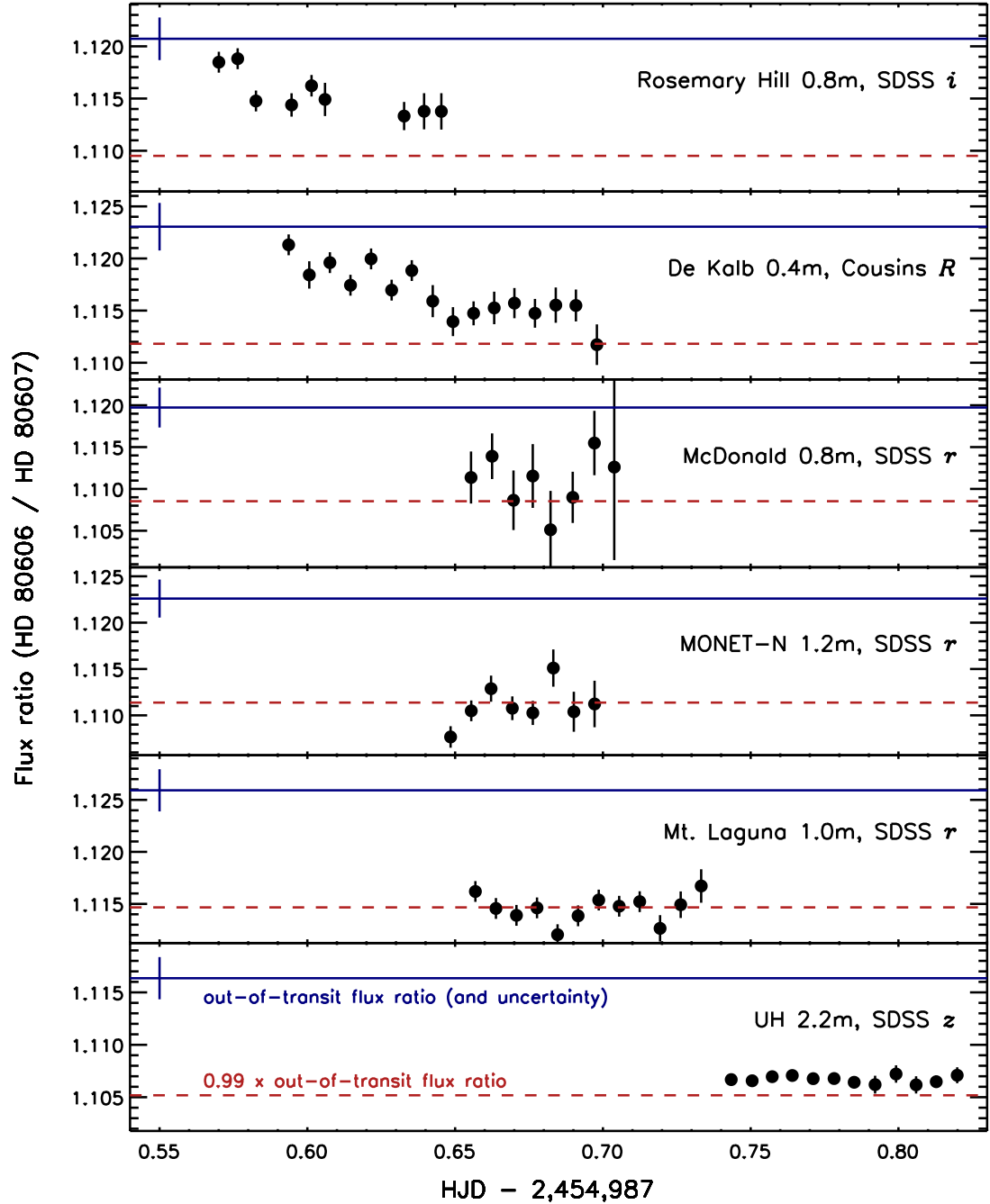}}
\end{center}
\vspace{-0.3in}
\caption{\label{fig:timeseries}
\normalsize
The flux ratio between HD~80606 and HD~80607, as measured on the
transit night UT~2009~June~5. The solid blue line is the
out-of-transit flux ratio as determined on a different night. The
uncertainty in the out-of-transit flux ratio is indicated with an
error bar on the left side. The dashed red line shows the 1\% drop
that was expected at midtransit.
}
\end{figure*}

%%%%%%%%%%%%%%%%%%%%%%%%%%%%%%%%%%%%%%%%%%%%%%%%%%%%%%%%%%%%%%%%%

Determining the out-of-transit flux ratio and its uncertainty was an
important task. Since it was not possible to gather out-of-transit
data on June~5, we needed to compare data from the same telescope that
were taken on different nights. Systematic errors are expected from
night-to-night differences in atmospheric conditions and detector
calibrations. We believe this uncertainty to be approximately 0.002 in
the flux ratio, based on the following two tests.

First, in two instances the out-of-transit flux ratio was measured on
more than one night, with differences in the results of 0.0004 and
0.0017 from the MONET-North and FLWO telescopes respectively. In the
latter case, the data were separated in time by 112~days, raising the
possibility that longer-term instabilities in the instrument or the
intrinsic variability of the stars contribute to the difference, and
that the night-to-night repeatability is even better than 0.0017.

A second comparison can be made by including data from {\it different}
telescopes that employed the same nominal bandpass. This should give
an upper bound (worst-case) estimate for the systematic error in the
measurement from a {\it single} telescope on different nights. Using
the data in Table~\ref{tbl:ratio}, we asked for each bandpass: what
value of $\sigma_{\rm sys}$ must be chosen in order for
\begin{equation}
\chi^2 =
\sum_{i=1}^{N} \frac{(f_{{\rm oot},i} - \bar{f}_{{\rm oot},i})^2}{\sigma_i^2 + \sigma_{\rm sys}^2 } = N-4,
\label{eq:chi2-offset}
\end{equation}
where $f_{{\rm oot},i}$ is the $i$th measurement of the out-of-transit
flux ratio, $\sigma_i$ is the statistical uncertainty in that
measurement, and $\bar{f}_{{\rm oot},i}$ is the unweighted mean of all
the out-of-transit flux ratio measurements made in the same nominal
bandpass as $f_i$. There are $N=15$ data points, and $N-4$ is used in
Eq.~(\ref{eq:chi2-offset}) because there are 4 independent bandpasses
for which means are calculated.\footnote{For this exercise we
  considered the ``narrow $z$'' band of the UH~2.2m observations to be
  equivalent to the Sloan $z$ band.} In this sense, we fitted a model
to the out-of-transit flux-ratio data with 4 free parameters. The
quantities $(f_{{\rm oot},i} - \bar{f}_{{\rm oot},i})$ are plotted in
Fig.~2. The result is $\sigma_{\rm sys}=0.0018$.

In our subsequent analysis we assumed the error in the out-of-transit
flux ratio to follow a Gaussian distribution with a standard deviation
given by the quadrature sum of 0.0020 and statistical error given in
Table~\ref{tbl:ratio}. Given the preceding results, we believe this to
be a reasonable and even a conservative estimate of the systematic
error. Though it may seem too small to those readers with experience
in synoptic photometry, it must be remembered that this is an
unusually favorable case: the Universe was kind enough to provide two
stars of nearly equal brightness and color separated by only
$20\arcsec$. It is also worth repeating that for our analysis we did
not need to place data from {\it different} telescopes on the same
flux scale; we needed only to align data from the {\it same} telescope
obtained on different nights.

\begin{figure}[t]
\epsscale{1.00}
\plotone{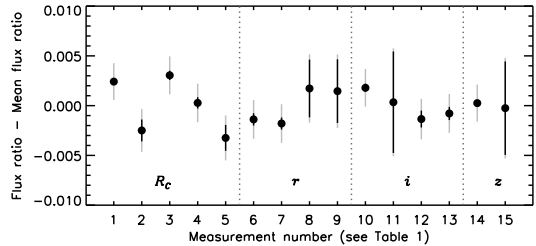}
\caption{ 
Deviations between the measured out-of-transit flux ratio, and the
mean value of the out-of-transit flux ratio across all data sets obtained
with the same nominal bandpass. The data are given in Table~\ref{tbl:ratio}. The black
error bars represent statistical errors. The gray error bars have an
additional systematic error of 0.0018 added in quadrature with the
statistical error. The value of 0.0018 was chosen because it gives a
reduced $\chi^2$ of unity [see Eq.~(\ref{eq:chi2-offset})].
\label{fig:offset}}
\end{figure}

We call attention to a few key aspects of the time series in
Fig.~\ref{fig:timeseries}: (1) All the observers measured the flux
ratio between HD~80606 and HD~80607 to be smaller on the transit night
than it was on out-of-transit nights. We conclude that the transit was
detected. (2) The data from Rosemary Hill and De Kalb show a decline
in the relative brightness of HD~80606 over several hours. We
interpret the decline as the transit ingress. (3) The data from
McDonald, Mt.\ Laguna, and Mauna Kea show little variability over the
interval of their observations, suggesting that the ``bottom''
(complete phase) of the transit had been reached.

\subsection{Radial Velocities}
\label{subsec:rv}

We measured the relative radial velocity (RV) of HD~80606 using the
Keck~I 10m telescope on Mauna Kea, Hawaii. We used the High Resolution
Echelle Spectrometer (HIRES; Vogt et al.~1994) in the standard setup
of the California Planet Search program (Howard et al.~2009), as
summarized here. We employed the red cross-disperser and used the
iodine gas absorption cell to calibrate the instrumental response and
the wavelength scale. The slit width was $0\farcs 86$ and the exposure
time ranged from 240--500~s, giving a resolution of $65,000$ and a
typical signal-to-noise ratio (SNR) of 210~pixel$^{-1}$. Radial
velocities were measured with respect to an iodine-free spectrum,
using the algorithm of Butler et al.~(1996) as improved over the
years.

The 73 measurements span 8~yr, from 2001 to the
present. Table~\ref{tbl:rv} gives all of the RV data. There are 39
data points obtained prior to the upgrade of the HIRES CCDs in August
2004, and 34 data points obtained after the upgrade. Results from the
pre-upgrade data, and some of the post-upgrade data, were published by
Butler et al.~(2006). For our analysis we re-reduced the post-upgrade
spectra using later versions of the analysis code and spectral
template. Due to known difficulties in comparing data obtained with
the different detectors, we allowed for a constant velocity offset
between the pre-upgrade and post-upgrade data sets in our subsequent
analysis.

\begin{figure*}[t]
\begin{center}
  \leavevmode
\hbox{%
  \epsfxsize=6in
  \epsffile{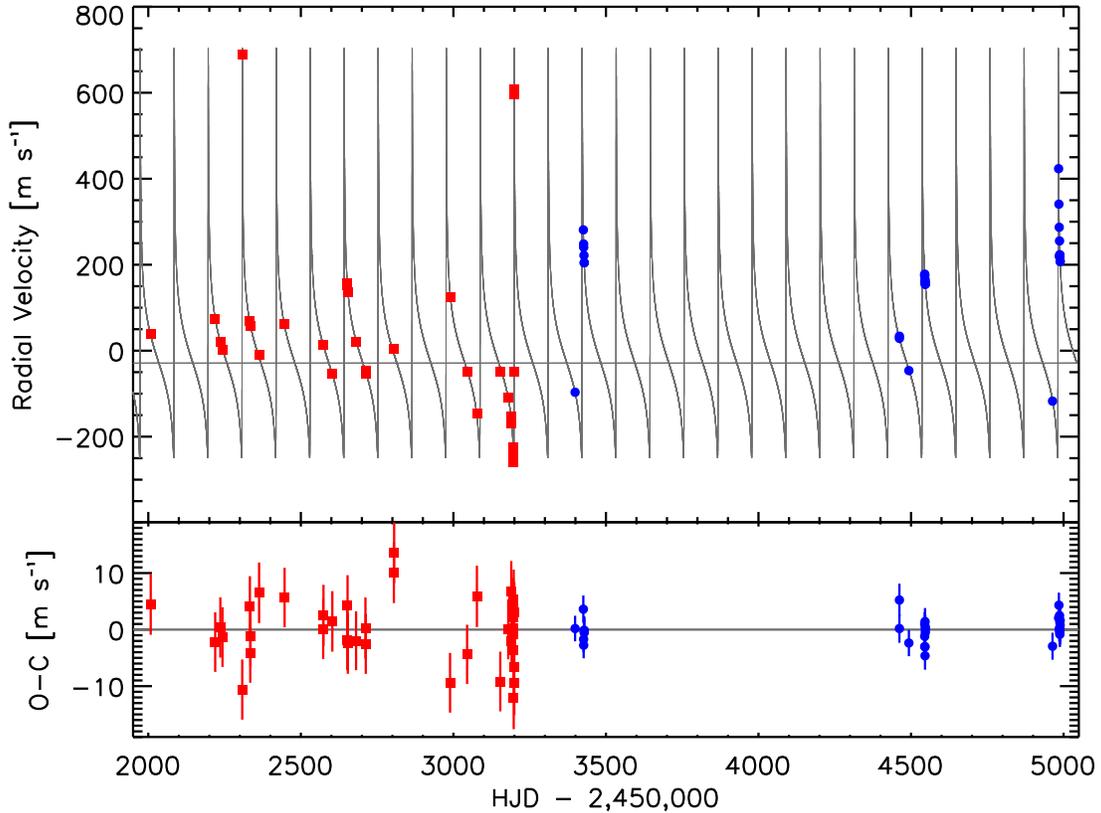}}
\end{center}
\caption{ Radial-velocity variation of HD~80606, as a function of time. Red
squares are the data obtained prior to the upgrade of the HIRES CCDs.
Blue dots are the post-upgrade data. The gray line is the
best-fitting model. The data were transformed into the center-of-mass
frame of the star-planet system by subtracting
a constant velocity offset, and the error bars
represent the quadrature sum of the measurement errors
quoted in Table~\ref{tbl:rv} and a term representing possible systematic
errors (``stellar jitter''). For the pre-upgrade and post-upgrade data, the constant velocity
offsets are $184.55$ and $182.45$~m~s$^{-1}$, and the systematic error
terms are 5 and 2~m~s$^{-1}$, respectively.
\label{fig:rv-time}}
\end{figure*}

\begin{figure*}[t]
\begin{center}
  \leavevmode
\hbox{%
  \epsfxsize=6in
  \epsffile{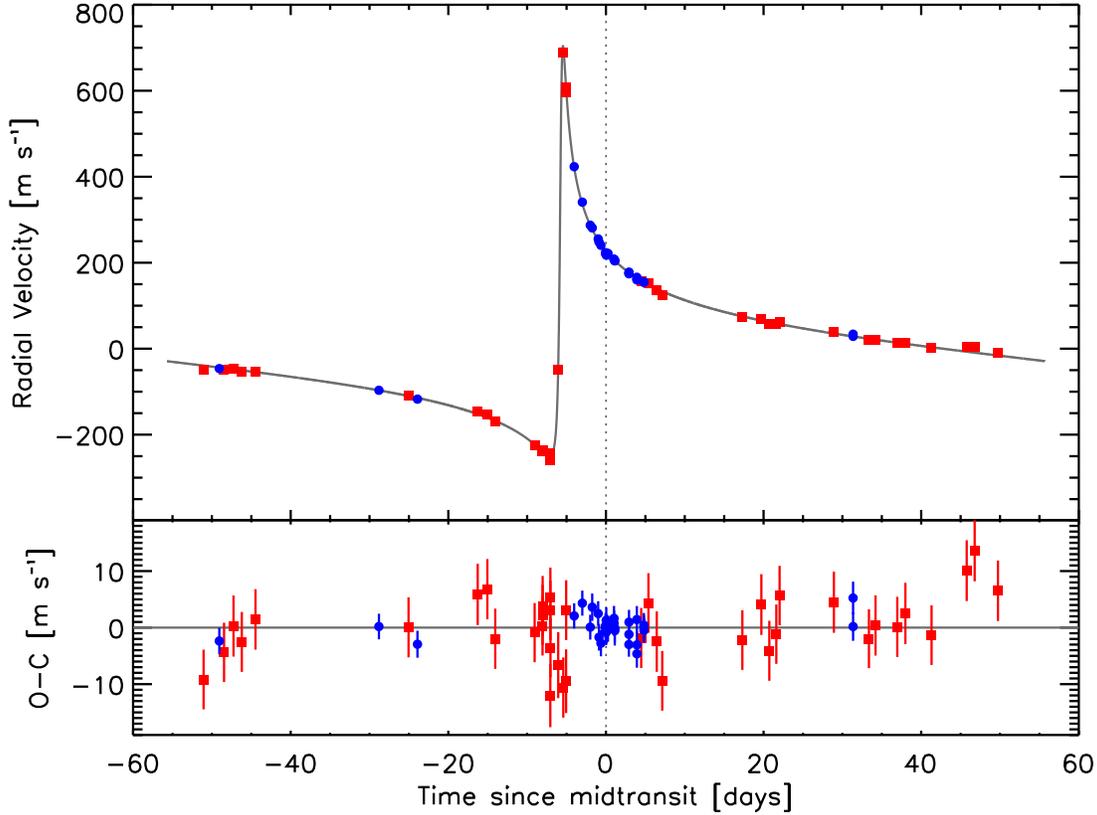}}
\end{center}
\caption{\label{fig:rv-phase}
\normalsize
Radial-velocity variation of HD~80606, as a function of orbital phase. The same
plotting conventions apply as in Fig.~\ref{fig:rv-time}.
}
\end{figure*}

\begin{figure*}[t]
\begin{center}
  \leavevmode
\hbox{%
  \epsfxsize=6in
  \epsffile{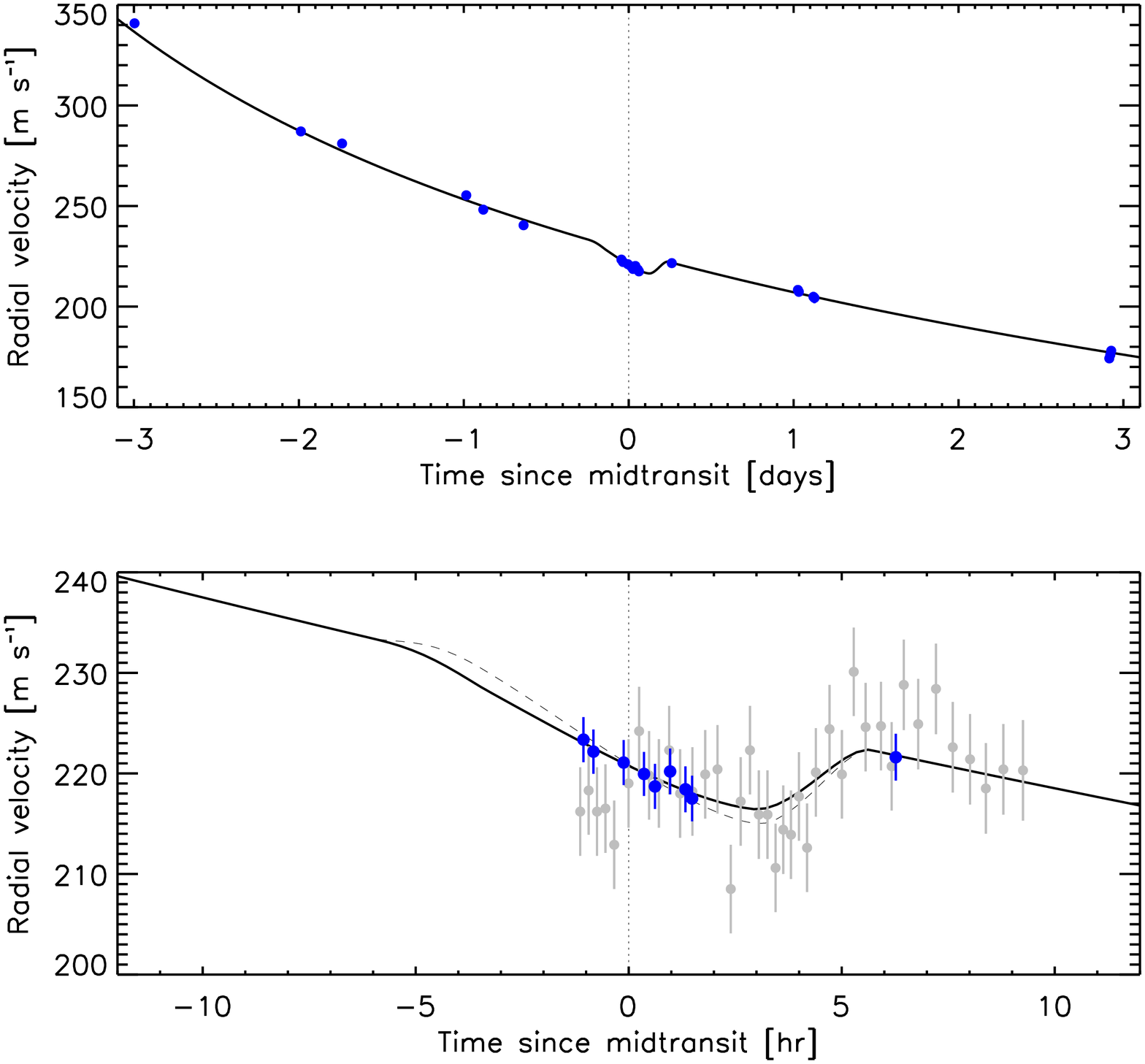}}
\end{center}
\caption{\label{fig:rv-transit}
\normalsize
Radial-velocity variation of HD~80606, as a function of orbital phase,
for the week of the transit (top panel) and the day of the transit
(bottom panel).
The in-transit RVs are all from 2009~June~5.
Of the out-of-transit RVs, 5 are from
the week of 2009~June~1--6, and the others are from different orbits.
Blue dots are the post-upgrade Keck/HIRES data, after
subtracting offsets and enlarging the error bars as in Figs.~\ref{fig:rv-phase}
and \ref{fig:rv-time}.
Gray dots are the SOPHIE data of Moutou et al.~(2009), which were not
used to derive the best-fitting models plotted here.
The solid line is the best-fitting model with no prior
constraint on $v\sin i_\star$. The dashed line is the best-fitting
model with a prior constraint on $v\sin i_\star$ as explained
in \S~\ref{sec:analysis}.
}
\end{figure*}

The post-upgrade data include nightly data from the week of the June~5
transit, which in turn include a series of 8 observations taken at
30~min intervals on the transit night. Fig.~\ref{fig:rv-time} shows
the RV data as a function of time, and Fig.~\ref{fig:rv-phase} shows
the RV data as a function of orbital phase. Fig.~\ref{fig:rv-transit}
is a close-up around the transit phase. Shown in all of these figures
is the best-fitting model, described in \S~3.

\section{Analysis}
\label{sec:analysis}

We fitted a model to the photometric and RV data based on the premise
of a single planet in a Keplerian orbit around a star with a quadratic
limb-darkening law and uniform rotation of its photosphere. The model
flux was computed using the equations of Mandel \& Agol (2002). The
model RV was given by $v_O(t) + \Delta v_R(t)$, where $v_O$ is the
line-of-sight component of the Keplerian orbital velocity and $\Delta
v_R$ is the anomalous velocity due to the Rossiter-McLaughlin (RM)
effect.

To compute $\Delta v_R$ as a function of orbital phase we used the
``RM calibration'' procedure of Winn et al.~(2005): we simulated
spectra exhibiting the RM effect at various orbital phases, and then
measured the apparent radial velocity of the simulated spectra using
the same algorithm used on the actual data. We found the results to be
consistent with the simple formula $\Delta v_R = -(\Delta f)v_p$ (Ohta
et al.~2005, Gim\'enez 2006), where $\Delta f$ is the instantaneous
decline in relative flux and $v_p$ is the radial velocity of the
hidden portion of the photosphere.\footnote{We also found this to be
  true for the cases of HAT-P-1 (Johnson et al.~2008) and TrES-2 (Winn
  et al.~2008a), although for other cases a higher-order polynomial
  relation was needed. It is noteworthy that the 3 systems for which
  the linear relation is adequate are the slowest rotators. This is
  consistent with work by T.\ Hirano et al.\ (in preparation) that
  aims at an analytic understanding of the RM calibration procedure.}

The model parameters can be divided into 3 groups. First are the
parameters of the spectroscopic orbit: the period $P$, a particular
midtransit time $T_t$, the radial-velocity semiamplitude $K$, the
eccentricity $e$, the argument of pericenter $\omega$, and two
velocity offsets $\gamma_1$ and $\gamma_2$ (for the pre-upgrade and
post-upgrade data). Next are the photometric parameters: the
planet-to-star radius ratio $R_p/R_\star$, the orbital inclination
$i$, the scaled stellar radius $R_\star/a$ (where $a$ is the semimajor
axis), and the out-of-transit flux ratio $f_{{\rm oot},i}$ specific to
the data from each telescope. Finally there are the parameters
relevant to the RM effect: the projected stellar rotation rate $v\sin
i_\star$ and the angle $\lambda$ between the sky projections of the
orbital axis and the stellar rotation axis [for illustrations of the
geometry, see Ohta et al.~(2005), Gaudi \& Winn (2007), or Fabrycky \&
Winn (2009)]. The limb-darkening (LD) coefficients were taken from the
tables of Claret~(2000, 2004), as appropriate for the bandpass of each
data set.\footnote{For the $R_C$ band, we used $u_1=0.3915$ and
  $u_2=0.2976$; for the $r$ band, $u_1=0.4205$ and $u_2=0.2911$; for
  the $i$ band, $u_1=0.3160$ and $u_2=0.3111$; and for the ``narrow
  $z$'' band, $u_1=0.2424$ and $u_2=0.3188$. We did not allow the LD
  coefficients to be free parameters because the photometric data are
  not precise enough to give meaningful constraints on them (and
  conversely, even large errors in the theoretical LD coefficients
  have little effect on our results).}

We fitted all the Keck/HIRES RV data and all the new photometric data
except the data from McDonald Observatory, which were the noisiest
data and gave redundant time coverage. To complete the phase coverage
of the transit, we also fitted the egress data of Fossey et al.~(2009)
obtained with the Celestron 0.35m telescope, which were the most
precise and exhibited the smallest degree of correlated noise.

The fitting statistic was a combination of the usual chi-squared
statistic and terms representing Gaussian {\it a priori} constraints.
Schematically,
\begin{equation}
\chi^2 = \chi^2_f + \chi^2_v + \chi^2_{\rm oot} + \chi^2_{\rm occ},
\label{eq:chi2}
\end{equation}
with the various terms defined as
\begin{eqnarray}
\chi^2_f & = & \sum_{i=1}^{N_f}
\left[ \frac{f_i({\mathrm{obs}}) - f_i({\mathrm{calc}})}{\sigma_{f,i}} \right]^2, \\
\chi^2_v & = & \sum_{i=1}^{N_v}
\left[ \frac{v_i({\mathrm{obs}}) - v_i({\mathrm{calc}})}{\sigma_{v,i}} \right]^2, \\
\chi^2_{\rm oot} & = & \sum_{i=1}^{4}
\left[ \frac{f_{\rm oot, i} - \bar{f}_{{\rm oot},i}}{0.0020} \right]^2, \\
\chi^2_{\rm occ} & = & 
\left[ \frac{T_o({\mathrm{obs}}) - T_o({\mathrm{calc}})}{\sigma_{T_o}} \right]^2 +
\left[ \frac{\tau_o({\mathrm{obs}}) - \tau_o({\mathrm{calc}})}{\sigma_{\tau_o}} \right]^2, \\
\end{eqnarray}
in which $f_i$(obs) is a measurement of the relative flux of HD~80606,
$\sigma_{f,i}$ is the uncertainty, and $f_i$(calc) is the relative
flux that is calculated for that time for a given set of model
parameters. Likewise $v_i$(obs) and $\sigma_{v,i}$ are the RV
measurements and uncertainties, and $v_i$(calc) is the calculated
RV. The third term enforces the constraints on the out-of-transit flux
ratios for each bandpass. The fourth term enforces constraints based
on the measured time and duration of the occultation; we adopt the
values $T_o=2,454,424.736\pm 0.004$~[HJD] and $\tau_o = 1.80\pm
0.25$~hr from Laughlin et al.~(2009). In contrast to previous analyses
(Pont et al.~2009), we did not impose prior constraints based on
theoretical stellar-evolutionary models, or on the stellar rotation
rate.  (In \S~\ref{subsec:spinorbit} we discuss how the results change
if such constraints are imposed.)

For the RV uncertainties $\sigma_{v,i}$, we used the quadrature sum of
the estimated measurement errors quoted in Table~\ref{tbl:rv}, and a
term $\sigma_{v, {\rm sys}}$ representing possible systematic
errors. The latter term is often called ``stellar jitter'' and may
represent Doppler shifts due to additional planets, non-Keplerian
Doppler shifts due to stellar oscillations or stellar activity, as
well as any errors in the instrument calibration or spectral
deconvolution code. We used $\sigma_{v, {\rm sys}}=5$~m~s$^{-1}$ for
the pre-upgrade data, and $\sigma_{v, {\rm sys}}=2$~m~s$^{-1}$ for the
post-upgrade data, based on the scatter in the observed RVs for other
planet-search program stars with similar spectral types that do not
have any detected planets.

With these choices, and with the flux uncertainties determined as
described previously, the minimum $\chi^2$ is 206 with 202 degrees of
freedom. This indicates a good fit and suggests that the estimated
uncertainties are reasonable. The rms scatter in the RV residuals is
5.7~m~s$^{-1}$ for the pre-upgrade data and 2.1~m~s$^{-1}$ for the
post-upgrade data. The rms scatter in the photometric residuals is
(respectively) 0.0015, 0.0012, 0.0013, and 0.00031 for the Rosemary
Hill, De Kalb, Mt.\ Laguna, and UH~2.2m data.

We determined the best fitting values of the model parameters and
their uncertainties using a Markov Chain Monte Carlo algorithm [see,
e.g., Tegmark et al.~(2004), Gregory~(2005), or Ford~(2005)]. This
algorithm creates a chain of points in parameter space by iterating a
jump function, which in our case was the addition of a Gaussian random
deviate to a randomly-selected single parameter. If the new point has
a lower $\chi^2$ than the previous point, the jump is executed; if
not, the jump is executed with probability $\exp(-\Delta\chi^2/2)$ and
otherwise the current point is repeated in the chain. We set the sizes
of the random deviates such that $\sim$40\% of jumps are executed. We
created 10 chains of $10^6$ links each from different starting
conditions, giving for each parameter a smoothly varying {\it a
  posteriori} distribution and a Gelman \& Rubin (1992) statistic
smaller than 1.05. The phase-space density of points in the chain is
an estimate of the joint {\it a posteriori}\, probability distribution
of all the parameters, from which may be calculated the probability
distribution for an individual parameter by marginalizing over all of
the others.

\vspace{0.5in}

\section{Results}

\begin{figure*}[t]
\begin{center}
  \leavevmode
\hbox{%
  \epsfxsize=6in
  \epsffile{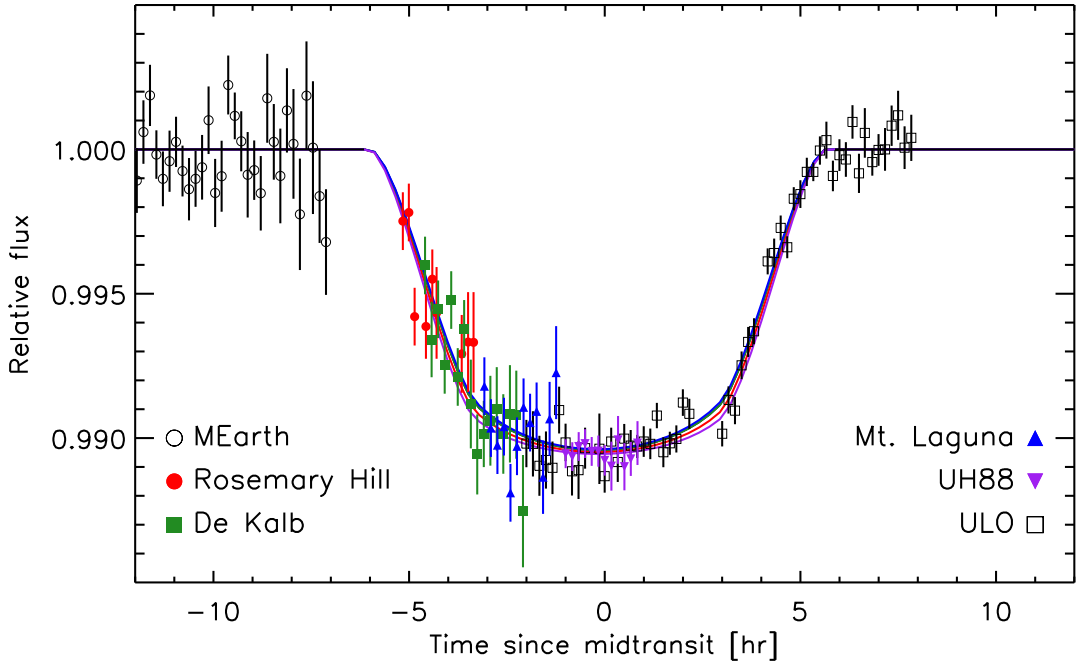}}
\end{center}
\caption{\label{fig:transit-lc}
\normalsize
The photometric transit of HD~80606. The solid curves
show the best-fitting model, which depends
on bandpass due to limb darkening. From top to bottom
the model curves are for the $r$, $R_C$, $i$, and $z$ bands.
}
\end{figure*}

Table~\ref{tbl:params} gives the results for the model parameters. The
quoted value for each parameter is the median of the {\it a
  posteriori}\, distribution, marginalized over all other
parameters. The quoted uncertainties represent 68.3\% confidence
limits, defined by the 15.85\% and 84.15\% levels of the cumulative
distribution. Fig.~\ref{fig:transit-lc} shows the transit light curve
and the best-fitting model. This figure also includes the the MEarth
observations of the 2009~Feb.~14 transit (Pont et al.~2009), which are
the most constraining of the available pre-ingress data.

\subsection{Spin-orbit parameters}
\label{subsec:spinorbit}

\begin{figure*}[t]
\begin{center}
  \leavevmode
\hbox{%
  \epsfxsize=6in
  \epsffile{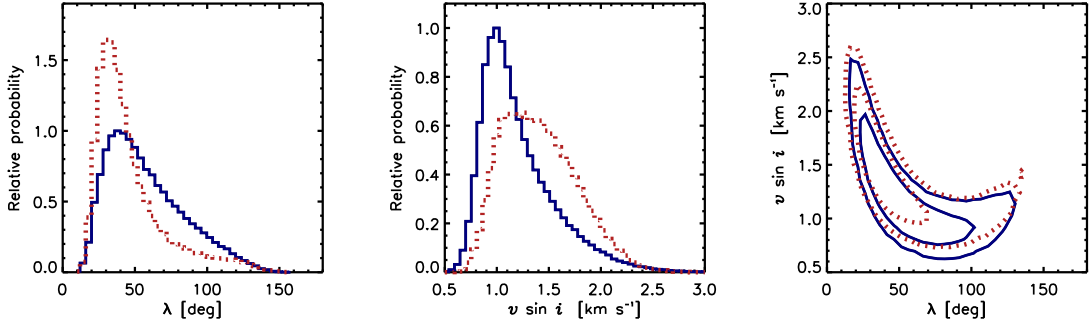}}
\end{center}
\vspace{-0.2in}
\caption{\label{fig:probdist} \normalsize Probability distributions
  for the projected spin-orbit angle ($\lambda$) and projected stellar
  rotation rate ($v\sin i_\star$).  Blue solid curves show the results
  when fitting the photometry and the Keck/HIRES RVs with no prior
  constraint on $v\sin i_\star$. Red dotted curves show the effect of
  applying a Gaussian prior $v\sin i_\star = 1.9\pm 0.5$~km~s$^{-1}$
  based on analyses of the stellar absorption lines in Keck/HIRES
  spectra.  {\it Left}.---Probability distribution for $\lambda$.
  {\it Center}.---Probability distribution for $v\sin i_\star$. {\it
    Right.}---Joint probability distribution for $v\sin i_\star$ and
  $\lambda$. The contours are the 68.3\% and 95\% confidence levels.
}
\end{figure*}

Fig.~\ref{fig:probdist} shows the probability distributions for the
parameters describing the Rossiter-McLaughlin effect, $v\sin i_\star$
and $\lambda$. A well-aligned system, $\lambda=0$, can be excluded
with high confidence.  With 68.3\% confidence, $\lambda$ lies between
32 and 87~deg, and with 99.73\% confidence, it lies between 14 and
142~deg. The distribution is non-Gaussian because of the correlation
between $\lambda$ and $v \sin i_\star$ shown in the right panel of
Fig.~\ref{fig:probdist}. No other parameter shows a significant
correlation with $\lambda$.

The strong exclusion of good alignment ($\lambda=0$) follows from the
observation that the RV data gathered on June~5 were blueshifted
relative to the Keplerian velocity (see Fig.~\ref{fig:rv-transit}),
over a time range that proved to include the midtransit time. Were the
spin and orbit aligned, the anomalous RV would vanish at midtransit,
because the planet would then be in front of the stellar rotation axis
where there is no radial component to the stellar rotation
velocity. The observed blueshift at midtransit implies that the
midpoint of the transit chord is on the redshifted (receding) side of
the star. This can only happen if the stellar rotation axis is tilted
with respect to the orbital axis.

For the projected stellar rotation rate, we find $v\sin i_\star=
1.12_{-0.22}^{+0.44}$~km~s$^{-1}$. In their previous analyses using
the SOPHIE data, Pont et al.~(2009) imposed prior constraints on
$v\sin i_\star$ based on the observed broadening in the stellar
absorption lines. This was necessary to break a degeneracy between the
transit duration, $v\sin i_\star$, and $\lambda$. Here we have
determined $v\sin i_\star$ by fitting the data exhibiting the RM
effect. This is preferable whenever possible, because of systematic
effects in both methods of determining $v\sin i_\star$. The method
based on the RM effect is subject to systematic errors in the ``RM
calibration'' procedure (see \S~\ref{sec:analysis} and Triaud et
al.~2009). The method based on line broadening is subject to
systematic error because of the degeneracy between rotational
broadening and other broadening mechanisms such as macroturbulence,
which are of comparable or greater magnitude for slowly rotating stars
such as HD~80606. These systematic effects may cause a mismatch in the
results for $v\sin i_\star$ obtained by the two methods, which could
lead to biased results for $\lambda$ if the result from the
line-broadening method were applied as a constraint on the RM model.

For comparison we review the spectroscopic determinations of $v\sin
i_\star$. Naef et al.~(2001) found $0.9\pm 0.6$~km~s$^{-1}$, based on
the width of the cross-correlation function measured with the ELODIE
spectrograph, after subtracting the larger ``intrinsic width'' due to
macroturbulence and other broadening mechanisms that was estimated
using the empirical calibration of Queloz et al.~(1998). This result
might be considered tentative, given that Queloz et al.~(1998) only
claim their calibration to be accurate down to
1.5--2~km~s$^{-1}$. Valenti \& Fischer~(2005) found $v\sin i_\star =
1.8\pm 0.5$~km~s$^{-1}$ based on synthetic spectral fitting to the
pre-upgrade Keck spectra, and a particular assumed relationship
between effective temperature and macroturbulence (see their paper for
details). We used the same spectral model and macroturbulence
relationship to analyze one of the post-upgrade Keck spectra, finding
$v\sin i_\star=2.0\pm 0.5$~km~s$^{-1}$, in good agreement with Valenti
\& Fischer (2005) but not Naef et al.~(2001).

We investigated the effect of imposing an {\it a priori}\, constraint
on $v\sin i_\star$ by adding the following term to Eq.~(\ref{eq:chi2}):
\begin{equation}
\chi^2_{{\rm rot}} = \left[ \frac{v\sin i_\star - 1.9~{\mathrm{km~s}^{-1}}}
                                {0.5~{\mathrm{km~s}}^{-1}} \right]^2.
\end{equation}
After refitting, the results for the spin-orbit parameters were $v\sin
i_\star = 1.37_{-0.33}^{+0.41}$~km~s$^{-1}$ and
$\lambda=39_{-13}^{+28}$~deg. The best-fitting model is shown with a dashed
line in Fig.~\ref{fig:rv-transit}. The constraints on $\lambda$ are
tightened; the new credible interval is 25\% smaller than the credible
interval without the constraint. However, the improved precision does
not necessarily imply improved accuracy, given the uncertainties
mentioned previously regarding the RM calibration and other broadening
mechanisms besides rotation. For this reason we have emphasized the
results with no external constraint on $v\sin i_\star$, and provide
only those results in Table~\ref{tbl:params}.

The preceding results did not make use of the SOPHIE data that Moutou
et al.~(2009) obtained during the February 2009 transit.  By leaving
out the SOPHIE data we have provided as independent a determination of
$\lambda$ as possible.  Fig.~\ref{fig:rv-transit} shows that the
SOPHIE data seem to agree with the Keck data, and with the
best-fitting model constrained by the Keck data.  When the SOPHIE data
are fitted simultaneously with the Keck data (with no prior constraint
on $v\sin i_\star$), the results for $\lambda$ are sharpened to
$59_{-16}^{+21}$~deg at 68.3\% confidence, and $59_{-35}^{+63}$~deg at
99.73\% confidence.

\subsection{Other parameters and absolute dimensions}

Our orbital parameters are generally in agreement with those derived
previously. One exception is the argument of pericenter, for which our
result ($300.83\pm 0.15$~deg) is 2$\sigma$ away from the result of
Laughlin et al.~(2009) ($300.4977\pm 0.0045$~deg), although the
uncertainty in the latter quantity seems likely to be
underestimated. Another exception is that our orbital period differs
from that of Laughlin et al.~(2009) by $3\sigma$, although our period
agrees with the period found by Pont et al.~(2009).

The transit parameters, including the transit duration, are related
directly to the stellar mean density $\rho_\star$ (Seager \&
Mallen-Ornelas~2003). In their previous studies, due to the poorly
known transit duration, Pont et al.~(2009) used theoretical
expectations for $\rho_\star$ to impose constraints on their
lightcurve solutions. Since we have measured the transit duration, we
can determine $\rho_\star$ directly from the data, finding $\rho_\star
= 1.63\pm 0.15$~g~cm$^{-3}$.\footnote{Although Seager \&
  Mallen-Ornelas~(2003) considered only circular orbits, their results
  are easily generalized. Needless to say we cannot assume a circular
  orbit in this case and our quoted uncertainty in $\rho_\star$
  incorporates the uncertainties in $e$ and $\omega$.}  This is
10--30\% larger than the Sun's mean density of $1.41$~g~cm$^{-3}$, as
expected for a metal-rich star with the observed G5 spectral type
(Naef et al.~2001).

\begin{figure}[ht]
\epsscale{1.0}
\plotone{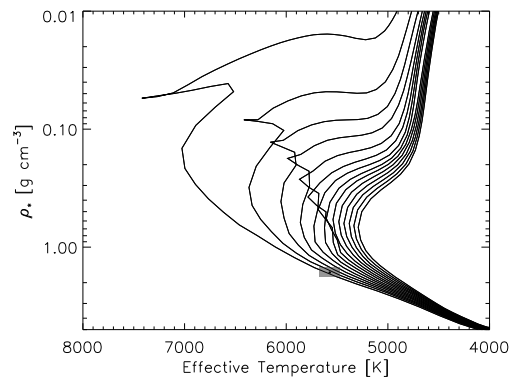}
\caption{ Stellar-evolutionary model isochrones in the space of
  effective temperature vs.\ stellar mean density, from the
  Yonsei-Yale series by Yi et al.\ (2001). The point and shaded box
  represent the observationally determined values and 68.3\%
  confidence intervals. Isochrones are shown for ages of 1 to 14 Gyr
  (from left to right) in steps of 1 Gyr for a fixed stellar
  metallicity of [Fe/H] = 0.344. \label{fig:isochrones}}
\end{figure}

We used this new empirical determination of $\rho_\star$ in
conjunction with stellar-evolutionary models to refine the estimates
of the stellar mass $M_\star$ and radius $R_\star$, which in turn lead
to refined planetary parameters (see, e.g., Sozzetti et al.~2007,
Holman et al.~2007). The models were based on the Yonsei-Yale series
(Yi et al.\ 2001; Demarque et al. 2004), and were applied as described
by Torres et al.~(2008) [with minor amendments by Carter et
al.~(2009)]. Figure~\ref{fig:isochrones} shows the theoretical
isochrones, along with some of the observational constraints. The
constraints were $\rho_\star=1.63\pm 0.15$~g~cm$^{-3}$, along with
$T_{\rm eff} = 5572\pm 100$~K and [Fe/H]~$= +0.34\pm 0.10$. The
temperature and metallicity estimates are based on the spectroscopic
analysis of Valenti \& Fischer (2005), but with enlarged error bars,
as per Torres et al.~(2008). The results are given in
Table~\ref{tbl:params}.

We did not apply any constraint to the models based on the
spectroscopically determined surface gravity ($\log g_\star$), out of
concern over systematic errors in that parameter (Winn et al.~2008b).
Instead we performed the reverse operation: given our results for
$M_\star$ and $R_\star$ we computed the implied value of $\log
g_\star$, finding $\log g_\star = 4.487\pm 0.021$. Reassuringly this
in agreement with, and is more precise than, the spectroscopically
determined values of $4.50\pm 0.20$ (Naef et al.~2001) and $4.44 \pm
0.08$ (Valenti \& Fischer 2005).

The RVs given in Table~\ref{tbl:rv} were measured with respect to an
arbitrary template spectrum, and therefore they do not represent the
true heliocentric RV of HD~80606, which is known much less
precisely. For reference, we measured the heliocentric RV to be
$\gamma = +3.95\pm 0.27$~km~s$^{-1}$, using the telluric ``A'' and
``B'' absorption bands of molecular oxygen to place the Keck/HIRES
spectra on the RV scale of Nidever et al.~(2002). This result is in
agreement with the previously reported value of $\gamma =
+3.767$~km~s$^{-1}$ (Naef et al.~2001).

\section{Summary and Discussion}

The poorly constrained transit duration was the main limiting factor
in previous determinations of the system parameters of HD~80606b. The
duration is now known to within 2.2\%, from a combination of the
transit ingress detected in our transcontinental campaign, the
photometric egress detected during the previous transit, and the
orbital period that is known very precisely from the RV data. In
addition, our new and more precise RV data show definitively that at
midtransit the starlight is anomalously blueshifted. This is
interpreted as the partial eclipse of the redshifted half of the
rotating photosphere. For this to happen at midtransit, the orbital
axis of the planet and the rotation axis of the star must be
misaligned.

Despite these achievements, the RV signal during the later phase of
the transit is known less precisely, and the RV signal during the
early phase of the transit remains unmeasured. This incompleteness
leads to relatively coarse bounds on the projected spin-orbit angle
$\lambda$ in comparison with many other systems.

As described in \S~1, the Kozai migration scenario of Wu \& Murray
(2003) carried an implicit prediction that the stellar spin and
planetary orbit are likely to be misaligned. In this sense, the
finding of a nonzero $\lambda$ corroborates the Kozai migration
hypothesis. The quantitative results for $\lambda$ derived in this
paper are in good agreement with the theoretical spin-orbit angle of
$50^\circ$ predicted by Fabrycky \& Tremaine~(2007) in an illustrative
calculation regarding HD~80606b (see their Fig.~1). This agreement
should not be overinterpreted, given the uncertainties in the
measurement, the issue of the sky projection, and the uncertainties in
some parameters of the calculation. Nevertheless the calculation
demonstrates that values of $\lambda$ of order $50^\circ$ emerge
naturally in the Kozai scenario.

The Kozai scenario is not without shortcomings. The orbital plane of
the stellar binary must be finely tuned to be nearly perpendicular to
the initial planetary orbit. This would be fatal to any scenario that
purported to explain the majority of exoplanetary orbits, but it may
be forgivable here, since we are trying to explain only one system out
of the several hundred known exoplanets. Another possible problem is
that (depending on the initial condition, and the characteristics of
the stellar binary) the relativistic precession may have been too
strong to permit Kozai oscillations (Naef et al.~2001). In this case
the theory might be rescued by the existence of a distant planet that
is responsible for the Kozai effect, rather than the stellar
companion. On the other hand, assuming HD~80607 is responsible,
additional planets would spoil the effect. Wu \& Murray~(2003) used
this fact to predict upper bounds on the masses and orbital distances
of any additional planets.

Another mechanism that can produce large eccentricities and large
spin-orbit misalignments is planet-planet scattering, in which close
encounters between planets cause sudden alterations in orbital
elements [Chatterjee et al.~2008, Juri{\'c} \& Tremaine~2008, Nagasawa
et al.~2008)]. However, Ford \& Rasio~(2008) found that planet-planet
scattering rarely produces eccentricities in excess of 0.8, unless the
orbit was initially eccentric (due perhaps to the Kozai effect, but
without the need for such extreme tuning) or the other planet that
participated in the encounter remained bound to the system.

For these reasons, further theoretical work is warranted, as is
continued RV monitoring to seek evidence for additional planets. On an
empirical level, it is striking that the only three exoplanetary
systems known to have a strong spin-orbit misalignment all have
massive planets on eccentric orbits: the present case of HD~80606b
(4.2~$M_{\rm Jup}$, $e=0.93$), WASP-14b (7.3~$M_{\rm Jup}$, $e=0.09$;
Joshi et al.~2009, Johnson et al.~2009), and XO-3b (11.8~$M_{\rm
  Jup}$, $e=0.26$; Johns-Krull et al.~2008, H\'ebrard et al.~2008,
Winn et al.~2009). There are also two cases of massive planets on
eccentric orbits for which $\lambda$ was found to be consistent with
zero: HD~17156b (3.2~$M_{\rm Jup}$, $e=0.68$; Cochran et al.~2008,
Barbieri et al.~2008, Narita et al.~2009) and HAT-P-2b (8.0~$M_{\rm
  Jup}$, $e=0.50$; Winn et al.~2007, Loeillet et al.~2008). Thus
although less massive planets on circular orbits seem to be
well-aligned, as a rule, it remains possible that more than half of
the massive eccentric systems are misaligned. Such systems are
fruitful targets for future RM observations.

\acknowledgments We thank Carly Chubak for measuring the heliocentric
radial velocity, Debra Fischer for help with measuring $v\sin
i_\star$, and Dan Fabrycky for interesting discussions about the Kozai
mechanism. We thank Philip Choi, Jason Eastman, Mark Everett, Scott
Gaudi, Zev Gurman, Marty Hidas, Matt Holman, and Alexander Rudy, for
their willingness to join this campaign even though they were not able
to participate due to weather or other factors. Mark Everett, Matt
Holman, and Dave Latham also helped to gather the out-of-transit data
from FLWO. We thank Bill Cochran and Ed Turner for help recruiting
participants. We also thank Steve Fossey, Jonathan Irwin, and David
Charbonneau for providing their data in a timely and convenient
manner. We are grateful to Greg Laughlin, whose enthusiasm and
persistence (and, it is suspected, a deal with the devil) led to the
discovery of the eclipses of HD~80606.

Some of the data presented herein were obtained at the W.M.~Keck
Observatory, which is operated as a scientific partnership among the
California Institute of Technology, the University of California, and
the National Aeronautics and Space Administration, and was made
possible by the generous financial support of the W.~M.~Keck
Foundation. We extend special thanks to those of Hawaiian ancestry on
whose sacred mountain of Mauna Kea we are privileged to be
guests. Without their generous hospitality, the Keck and UH~2.2m
observations presented herein would not have been possible. The MONET
network is funded by the Alfried Krupp von Bohlen und
Halbach-Stiftung. WFW and SK acknowledge support from NASA under grant
NNX08AR14G issued through the Kepler Discovery Program. KDC, EBF, FJR,
and observations from Rosemary Hill Observatory were supported by the
University of Florida. Observations at Mt. Laguna Observatory were
supported by the HPWREN network, funded by the National Science
Foundation Grant Numbers 0087344 and 0426879 to University of
California, San Diego. JAJ acknowledges support from an NSF Astronomy
and Astrophysics Postdoctoral Fellowship (grant no.\
AST-0702821). Work by JNW was supported by the NASA Origins program
through awards NNX09AD36G and NNX09AB33G. JNW also gratefully
acknowledges the support of the MIT Class of 1942 Career Development
Professorship.

{\it Facilities:} \facility{Keck:I (HIRES)} \facility{FLWO:1.2m
  (Keplercam)}, \facility{UH:2.2m (OPTIC)}, \facility{McD:0.8m}

%%%%%%%%%%%%%%%%%%%%%%%%%%%%%%%%%%%%%%%%%%%%%%%%%%%%%%%%%%%%%%%%%

\begin{deluxetable*}{lcc}

\tabletypesize{\scriptsize}
\tablecaption{Relative Radial Velocity Measurements of HD~80606\label{tbl:rv}}
\tablewidth{0pt}

\tablehead{
\colhead{HJD} &
\colhead{RV [m~s$^{-1}$]} &
\colhead{Error [m~s$^{-1}$]}
}

\startdata
  $  2452007.89717$  &  $   -144.75$  &  $   2.06$  \\
  $  2452219.16084$  &  $   -111.52$  &  $   1.68$  \\
  $  2452236.05808$  &  $   -163.28$  &  $   1.85$  \\
  $  2452243.16763$  &  $   -182.50$  &  $   1.72$  \\
  $  2452307.87799$  &  $    505.19$  &  $   1.82$  \\
  $  2452333.00899$  &  $   -114.90$  &  $   1.97$  \\
  $  2452334.01381$  &  $   -126.84$  &  $   1.78$  \\
  $  2452334.90457$  &  $   -127.12$  &  $   1.59$  \\
  $  2452363.02667$  &  $   -193.85$  &  $   1.91$  \\
  $  2452446.75474$  &  $   -121.75$  &  $   1.63$  \\
  $  2452573.12626$  &  $   -170.54$  &  $   1.83$  \\
  $  2452574.15413$  &  $   -170.69$  &  $   2.10$  \\
  $  2452603.13145$  &  $   -237.61$  &  $   1.91$  \\
  $  2452652.07324$  &  $    -27.86$  &  $   1.80$  \\
  $  2452652.99160$  &  $    -31.09$  &  $   1.67$  \\
  $  2452654.04838$  &  $    -47.65$  &  $   1.89$  \\
  $  2452680.96832$  &  $   -163.44$  &  $   1.43$  \\
  $  2452711.77421$  &  $   -232.28$  &  $   2.05$  \\
  $  2452712.83539$  &  $   -237.55$  &  $   1.76$  \\
  $  2452804.81344$  &  $   -181.39$  &  $   2.00$  \\
  $  2452805.81998$  &  $   -180.13$  &  $   2.12$  \\
  $  2452989.05542$  &  $    -60.43$  &  $   1.71$  \\
  $  2453044.89610$  &  $   -234.22$  &  $   1.60$  \\
  $  2453077.07929$  &  $   -330.76$  &  $   2.11$  \\
  $  2453153.73469$  &  $   -233.14$  &  $   1.70$  \\
  $  2453179.74129$  &  $   -294.54$  &  $   1.80$  \\
  $  2453189.74093$  &  $   -338.05$  &  $   1.85$  \\
  $  2453190.73876$  &  $   -354.28$  &  $   1.84$  \\
  $  2453195.73688$  &  $   -408.34$  &  $   1.55$  \\
  $  2453196.74363$  &  $   -423.32$  &  $   1.62$  \\
  $  2453196.75039$  &  $   -421.51$  &  $   1.77$  \\
  $  2453196.75701$  &  $   -420.03$  &  $   1.84$  \\
  $  2453197.73139$  &  $   -444.83$  &  $   2.54$  \\
  $  2453197.73809$  &  $   -427.39$  &  $   1.63$  \\
  $  2453197.74463$  &  $   -429.67$  &  $   1.67$  \\
  $  2453197.75126$  &  $   -436.21$  &  $   1.60$  \\
  $  2453198.73328$  &  $   -232.36$  &  $   3.02$  \\
  $  2453199.73285$  &  $    412.11$  &  $   2.61$  \\
  $  2453199.73960$  &  $    422.80$  &  $   1.76$  \\
  $  2453398.85308$  &  $   -279.18$  &  $   1.06$  \\
  $  2453425.92273$  &  $     98.61$  &  $   1.36$  \\
  $  2453426.77899$  &  $     65.74$  &  $   1.24$  \\
  $  2453427.02322$  &  $     57.99$  &  $   1.17$  \\
  $  2453427.92202$  &  $     39.16$  &  $   1.19$  \\
  $  2453428.78092$  &  $     22.40$  &  $   1.48$  \\
  $  2453428.78724$  &  $     21.85$  &  $   1.72$  \\
  $  2454461.95594$  &  $   -148.74$  &  $   2.09$  \\
  $  2454461.96114$  &  $   -153.79$  &  $   1.54$  \\
  $  2454492.97860$  &  $   -228.73$  &  $   1.20$  \\
  $  2454544.94958$  &  $     -8.21$  &  $   0.81$  \\
  $  2454544.95553$  &  $     -6.49$  &  $   0.84$  \\
  $  2454544.96119$  &  $     -4.40$  &  $   0.86$  \\
  $  2454545.95602$  &  $    -16.31$  &  $   1.32$  \\
  $  2454545.96140$  &  $    -22.42$  &  $   1.44$  \\
  $  2454545.96684$  &  $    -20.91$  &  $   1.28$  \\
  $  2454546.84875$  &  $    -27.09$  &  $   0.80$  \\
  $  2454546.85412$  &  $    -27.53$  &  $   0.77$  \\
  $  2454546.85975$  &  $    -28.14$  &  $   0.82$  \\
  $  2454963.87159$  &  $   -299.78$  &  $   1.30$  \\
  $  2454983.75577$  &  $    240.93$  &  $   0.99$  \\
  $  2454984.78842$  &  $    158.39$  &  $   0.97$  \\
  $  2454985.79701$  &  $    104.64$  &  $   0.92$  \\
  $  2454986.80004$  &  $     72.89$  &  $   0.88$  \\
  $  2454987.74080$  &  $     40.91$  &  $   1.00$  \\
  $  2454987.75055$  &  $     39.72$  &  $   0.92$  \\
  $  2454987.78009$  &  $     38.64$  &  $   1.00$  \\
  $  2454987.79998$  &  $     37.50$  &  $   0.88$  \\
  $  2454987.81073$  &  $     36.26$  &  $   1.01$  \\
  $  2454987.82554$  &  $     37.75$  &  $   1.08$  \\
  $  2454987.84052$  &  $     35.98$  &  $   1.09$  \\
  $  2454987.84704$  &  $     35.06$  &  $   1.07$  \\
  $  2454988.81104$  &  $     25.85$  &  $   0.92$  \\
  $  2454988.81684$  &  $     24.89$  &  $   1.00$
\enddata

\tablecomments{The RV was measured relative to an arbitrary template
  spectrum; only the differences are significant. The uncertainty
  given in Column 3 is the internal error only and does not account
  for any possible ``stellar jitter.''}

\end{deluxetable*}

%%%%%%%%%%%%%%%%%%%%%%%%%%%%%%%%%%%%%%%%%%%%%%%%%%%%%%%%%%%%%%%%%

\tabletypesize{\normalsize}

\begin{deluxetable}{lcc}

% \tabletypesize{\footnotesize}
\tablecaption{System Parameters of HD~80606\label{tbl:params}}
\tablewidth{0pt}

\tablehead{
\colhead{Parameter} &
\colhead{Value} &
\colhead{Uncertainty}
}

\startdata
Orbital period, $P$~[d]                            & $111.43740$ & $0.00072$ \\
Midtransit time~[HJD]                              & $2,454,987.7842$ & $0.0049$ \\
Transit duration~(first to fourth contact)~[hr]    &  $11.64$  &  $0.25$    \\
Transit ingress or egress duration~[hr]            &  $2.60$   &  $0.18$    \\
\hline
Midoccultation time~[HJD]                             & $2,454,424.736$ & $0.004$ \\
Time from occultation to transit [d]                 & $5.8585$ & $0.0079$ \\
Occultation duration~(first to fourth contact)~[hr]    &  $1.829$  &  $0.056$  \\
Occultation ingress or egress duration~[hr]            &  $0.1725$   &  $0.0063$    \\
\hline
Velocity semiamplitude, $K$~[m~s$^{-1}$]            & $476.1$    &  $2.2$ \\
Orbital eccentricity, $e$                          & $0.93286$ &  $0.00055$ \\
Argument of pericenter, $\omega$~[deg]             & $300.83$ &  $0.15$ \\
Velocity offset, pre-upgrade [m~s$^{-1}$]           & $-184.58$ &  $0.93$ \\
Velocity offset, post-upgrade [m~s$^{-1}$]          & $-182.46$ &  $0.66$ \\
Heliocentric velocity [km~s$^{-1}$]                 & $+3.95$ &  $0.27$ \\
\hline
Planet-to-star radius ratio, $R_p/R_\star$           & $0.1033$  &  $0.0011$ \\
Orbital inclination, $i$~[deg]                     & $89.324$  & $0.029$     \\
Scaled semimajor axis, $a/R_\star$                   &  $102.4$   &  $2.9$      \\
Semimajor axis, $a$~[AU]                            &  $0.4614$   &  $0.0047$  \\
Transit impact parameter                            &  $0.788$  &  $0.016$   \\
Occultation impact parameter                        &  $0.0870$  &  $0.0019$   \\
\hline
Projected stellar rotation rate, $v \sin i_\star$~[km~s$^{-1}$]   &  $1.12$ &  $-0.22$,$+0.44$ \\
Projected spin-orbit angle, $\lambda$~[deg]         &  $53$  & $-21$,$+34$ \\
\hline
{\it ~~~Stellar parameters:} \\
Mass, $M_\star$~[$M_\odot$]        &  $1.05$  & $0.032$ \\
Radius, $R_\star$~[$R_\odot$]        &  $0.968$  & $0.028$ \\
Luminosity\tablenotemark{b}, $L_\star$~[$L_\odot$]        &  $0.801$  & $0.087$ \\
Mean density, $\rho_\star$~[g~cm$^{-3}$]          &  $1.63$  & $0.15$ \\
Surface gravity, $\log g_\star$~[cm~s$^{-2}$] &  $4.487$  & $0.021$ \\
Effective temperature\tablenotemark{b}, $T_{\rm eff}$ [K] &  $5572$  & $100$ \\
Metallicity\tablenotemark{b}, [Fe/H] &  $0.34$  & $0.10$ \\
Age [Gyr]  &  $1.6$  & $-1.1$,$+1.8$ \\
Distance\tablenotemark{a} [pc]  &  $61.8$  & $3.8$ \\
\hline
{\it ~~~Planetary parameters:} \\
Mass, $M_p$~[$M_{\rm Jup}$]        &  $4.20$  & $0.11$ \\
Mass ratio, $M_p/M_\star$        &  $0.00382$  & $0.00016$ \\
Radius, $R_p$~[$R_{\rm Jup}$]        &  $0.974$  & $0.030$ \\
Mean density, $\rho_\star$~[g~cm$^{-3}$]           &  $5.65$  & $0.54$ \\
Surface gravity, $g_p$~[m~s$^{-2}$] & $110.5$ & $8.2$ \\
\hline
\enddata

\tablecomments{Based on the joint analysis of the Keck/HIRES RV data,
  our new photometric data, the Celestron data of Fossey et
  al.~(2009), and the occultation time and duration measured by
  Laughlin et al.~(2009), except where noted.}

\tablenotetext{a}{Based on the apparent $V$ magnitude of 9.06 and the
  luminosity implied by the stellar-evolutionary models.}

\tablenotetext{b}{Based on an analysis of the iodine-free Keck/HIRES
  spectrum using the {\it Spectroscopy Made Easy} (SME) spectral
  synthesis code; see Valenti \& Piskunov~(1996), Valenti \& Fischer
  (2005).}

\end{deluxetable}

%%%%%%%%%%%%%%%%%%%%%%%%%%%%%%%%%%%%%%%%%%%%%%%%%%%%%%%%%%%%%%%%%

\end{document}